\documentclass[fleqn,12pt,twoside]{article}
\usepackage{espcrc1}

\usepackage{graphicx}

\newcommand{\AmS}{{\protect\the\textfont2
  A\kern-.1667em\lower.5ex\hbox{M}\kern-.125emS}}

\title{Time-dependent wave-packet approach for
fusion reactions of halo nuclei}

\author{K. Yabana\address[Tsukuba]{Institute of Physics, University of
Tsukuba, Tsukuba 305-8571, Japan},
M. Ueda\addressmark[Tsukuba]
and
T. Nakatsukasa\address[Tohoku]{Physics Department, Tohoku University,
Sendai 980-8578, Japan}
}

\begin{document}

\maketitle

\begin{abstract}
The fusion reaction of a halo nucleus $^{11}$Be on $^{208}$Pb
is described by a three-body direct reaction model.
A time-dependent wave packet approach is applied to a
three-body reaction problem. The wave packet approach enables us
to obtain scattering solutions without considering the three-body
scattering boundary conditions. The time evolution of the wave 
packet also helps us to obtain intuitive understanding of the
reaction dynamics. The calculations indicate a decrease of the 
fusion probability by the presence of the halo neutron.
\end{abstract}

\section{INTRODUCTION}

Since the discovery of the halo nuclei, much effort has been devoted
to develop reaction theories appropriate for weakly bound projectiles.
The eikonal approximation has been found to be useful for reactions
at medium and high incident energies\cite{Yabana92}.
However, even a basic picture has not yet been established for the 
reactions at low incident energies. For example, there have been 
controversial arguments on whether the fusion probability is enhanced
or suppressed by the presence of the halo nucleon.

We have been analyzing the low energy reactions of halo nuclei in a
three-body direct reaction model. A time-dependent wave packet
approach has been developed for this purpose. We first analyzed a
one-dimensional three-body model\cite{Yabana95}. We then made analysis 
of the fusion reaction of $^{11}$Be on a medium mass target, $^{40}$Ca, 
in a three-dimensional model\cite{Yabana97}.  The calculation was 
restricted to the cases of total angular momentum being equal to zero 
(head-on collision). 
In these analyses, we have found that the fusion probability 
is hindered by the presence of the halo neutron. This conclusion is 
opposite to many other theoretical works which have claimed the increase
of the fusion probability\cite{Takigawa91,Hussein92}.

We here report our extended analyses with the wave packet method for
the reaction of $^{11}$Be on $^{208}$Pb. For this reaction, the analyses 
have been made recently with the three-body model by other 
groups\cite{Hagino00,Diaz-Torres02}. The measurements are also 
available for systems close to this\cite{Yoshida96}.

The Hamiltonian of the three-body model is time-independent. Still, there are
several reasons to employ the time-dependent approach for the static 
problem. The time evolution of the wave packet gives us intuitive 
pictures for the reaction mechanism. It is not necessary to prepare 
complicated scattering boundary conditions for the three-body final 
states. The reaction probabilities for a certain energy region can 
be obtained at once from a single wave-packet solution. The trade-off for
these advantages is a heavy computational cost. The method has also
been developed and applied in the field of chemical reactions\cite{Meijer98}.

\section{THE THREE-BODY MODEL}

We describe a reaction of a single-neutron halo nucleus by the 
three-body model consisting of the halo neutron (n), the core nucleus 
(C), and the target nucleus (T). The projectile (P) is composed of the 
halo neutron and the core nucleus, P=n+C. We consider the reaction
of $^{11}$Be on $^{208}$Pb. Namely, the core is the $^{10}$Be and the 
target is the $^{208}$Pb. The time-dependent Schr\"odinger equation 
is expressed as
\begin{equation}
i\hbar \frac{\partial}{\partial t} \psi({\bf R}, {\bf r}, t)
=
\left[ -\frac{\hbar^2}{2\mu} \nabla_{\bf R}^2
       -\frac{\hbar^2}{2m}   \nabla_{\bf r}^2
       +V_{nC}(r) +V_{CT}(R_{CT}) +V_{nT}(r_{nT})
\right] \psi({\bf R},{\bf r}, t),
\label{TDSE}
\end{equation}
where we denote the relative n-C coordinate as ${\bf r}$ and the
relative P-T coordinate as ${\bf R}$. The reduced masses of
n-C and P-T motions are $m$ and $\mu$, respectively.
The n-C potential, $V_{nC}(r)$, is real.
The n-T potential, $V_{nT}(r_{nT})$, is also
chosen to be real. The C-T potential, $V_{CT}(R_{CT})$, is complex.
The real part of $V_{CT}$ includes the Coulomb and nuclear potentials.
The imaginary part of $V_{CT}$ describes the fusion process as a loss
of the flux in the reaction between the core and the target nuclei.
All the nuclear potentials are taken to be of Woods-Saxon shape.
For the n-C potential, $r_{nC}=1.3$ fm and $a_{nC}=0.7$ fm. The strength
of the potential is set to give the $2s$ orbital energy at $-0.6$ MeV,
close to the neutron separation energy of the $^{11}$Be nucleus.
For the n-T potential, $r_{nT}=1.27$ fm and $a_{nT}=0.67$ fm. The strength
will be varied. For the real part of the C-T potential, we employ
$r_{CT}^R=1.192$ fm, $a_{CT}^R=0.63$ fm, and $V_{CT}=-46.764$ MeV.
The Coulomb potential is that of the uniformly charged sphere with the
radius parameter, $r_{CT}^C = 1.26$ fm. For the imaginary part of the
C-T potential, we employ $r_{CT}^I=0.8$ fm, $a_{CT}^I=0.4$ fm, and
$W_{CT}=-50$ MeV.

In this three-body model, we define the fusion process of the
three-body reaction as a loss of the flux caused by the absorptive 
C-T potential. Namely, the fusion is supposed to occur between 
the core and the target nuclei when they come closer beyond the Coulomb 
barrier. In this definition, the fusion probability includes both 
the complete and incomplete fusions, since the lost flux does not
distinguish final states of the neutron.

In practical calculations, we make a partial wave expansion.
There are two options for this; the body-fixed representation and the
space-fixed one\cite{Pack74}. For computations at finite total
angular momenta, the former is superior to the latter. Here we 
only consider the reactions with zero total angular momentum, for 
which both representations merge into the same expression,
\begin{equation}
\psi^{J=0}({\bf R},{\bf r},t)
=
\sum_{l=0}^{l_{max}} \frac{u_l(R,r,t)}{Rr} \frac{\sqrt{2l+1}}{4\pi} P_l(\cos\theta),
\end{equation}
where $\theta$ represents the angle between the two vectors 
${\bf R}$ and ${\bf r}$.
The $l$ is the relative angular momentum of the n-C motion, which is now
identical to that of the P-T motion in the case of $J=0$.
In numerical calculations, one has to restrict the summation within a
finite range, $l \leq l_{max}$.

As the initial wave packet, we employ a Gaussian packet for the
P-T motion,
\begin{equation}
u_0(R,r,t_0) = \exp[-\gamma(R-R_0)^2-iK_0 R] \phi_0(r) ,
\end{equation}
where $\phi_0(r)$ is the $2s$ orbital of the n-C motion and
$u_l=0$ for $l\neq 0$ at $t=t_0$.
The $R_0$ specifies the P-T separation at the center of the wave packet
and $K_0$ indicates the P-T wave number on average at $t=t_0$.
This wave packet state has the average total energy approximately given
by $E_0 = \hbar^2 K_0^2/2\mu + V_{CT}(R_0) - \epsilon_0$,
where $\epsilon_0$ is the binding energy of the orbital $\phi_0$.
The $R_0$ is set to be large so that $V_{CT}$ and $V_{nT}$ vanish except 
for the C-T Coulomb potential. 
We set $R_0=25$ fm and $\gamma=0.05$ fm$^{-2}$.

To solve the time-dependent Schr\"odinger equation for the radial
wave function $u_l(R,r,t)$, we discretize the radial variables.
We treat radial region $0<R<50$ fm and $0<r<60$ fm with radial step
$\Delta R=0.5$ fm and $\Delta r = 1$ fm, respectively. The discrete variables
representation is used for the second-order differential 
operator\cite{Colbert92}. It is important to employ very high
partial waves $l$ to describe the transfer reactions in the
Jacobi coordinate of the incident channel. We take up to $l_{max}=70$.
The wave function $u_l(R,r,t)$ is thus described on the grid points
of about 420,000.
For the time evolution, we employ the Taylor expansion method.
Previously we employed the Crank-Nicholson formula\cite{Yabana97}.
The Taylor expansion method requires usage of the smaller time step
and gives more accurate results. 
The time evolution is continued until most parts of the
wave packet leave the interaction region after scatterings and the
potentials from the target again becomes negligible except for the Coulomb.

\section{WAVE PACKET DYNAMICS}

\begin{figure}[tbp]
\begin{center}
\begin{tabular}{ccc}
\resizebox{!}{36mm}{\includegraphics{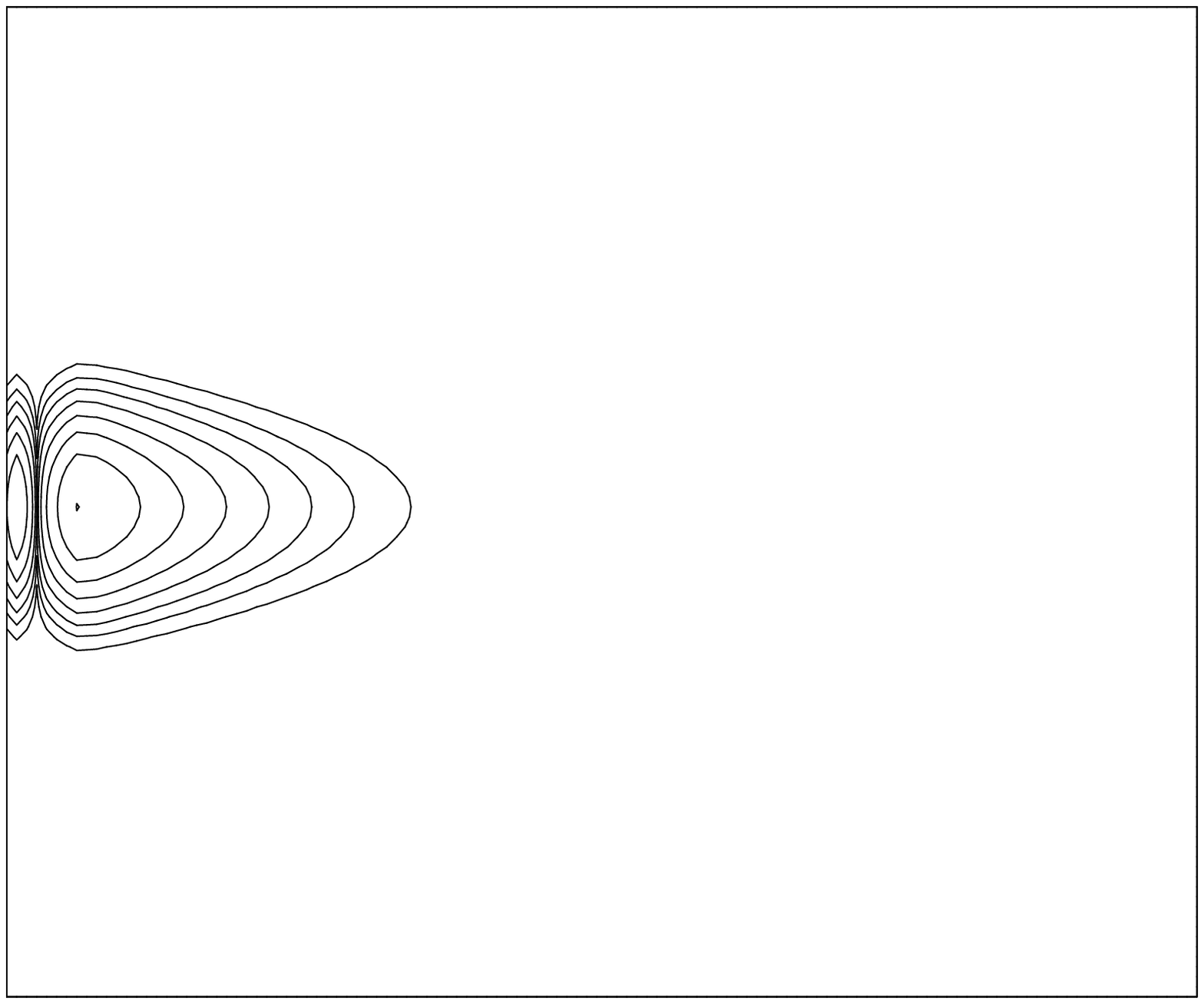}} & &
\resizebox{!}{36mm}{\includegraphics{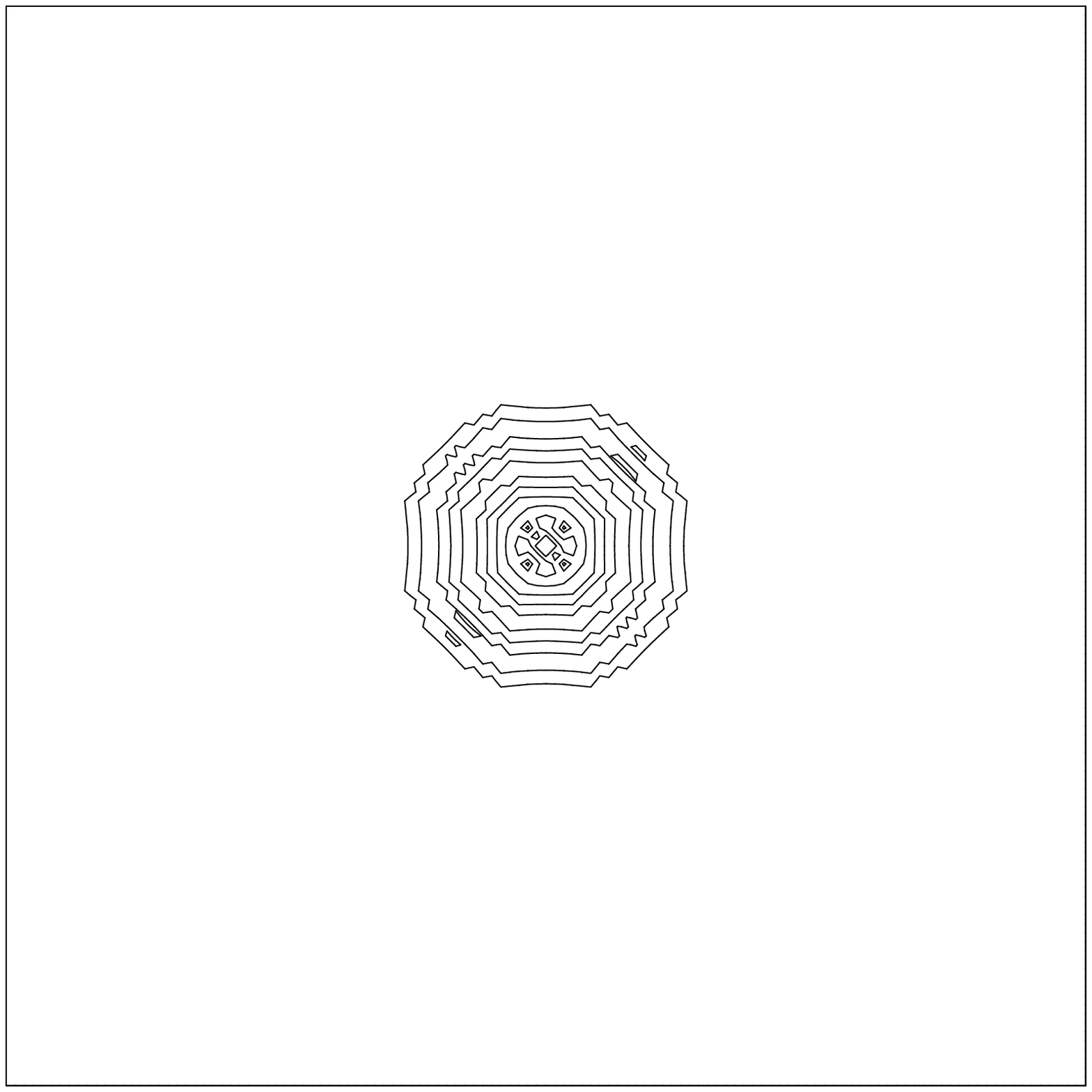}}  \\
\resizebox{!}{36mm}{\includegraphics{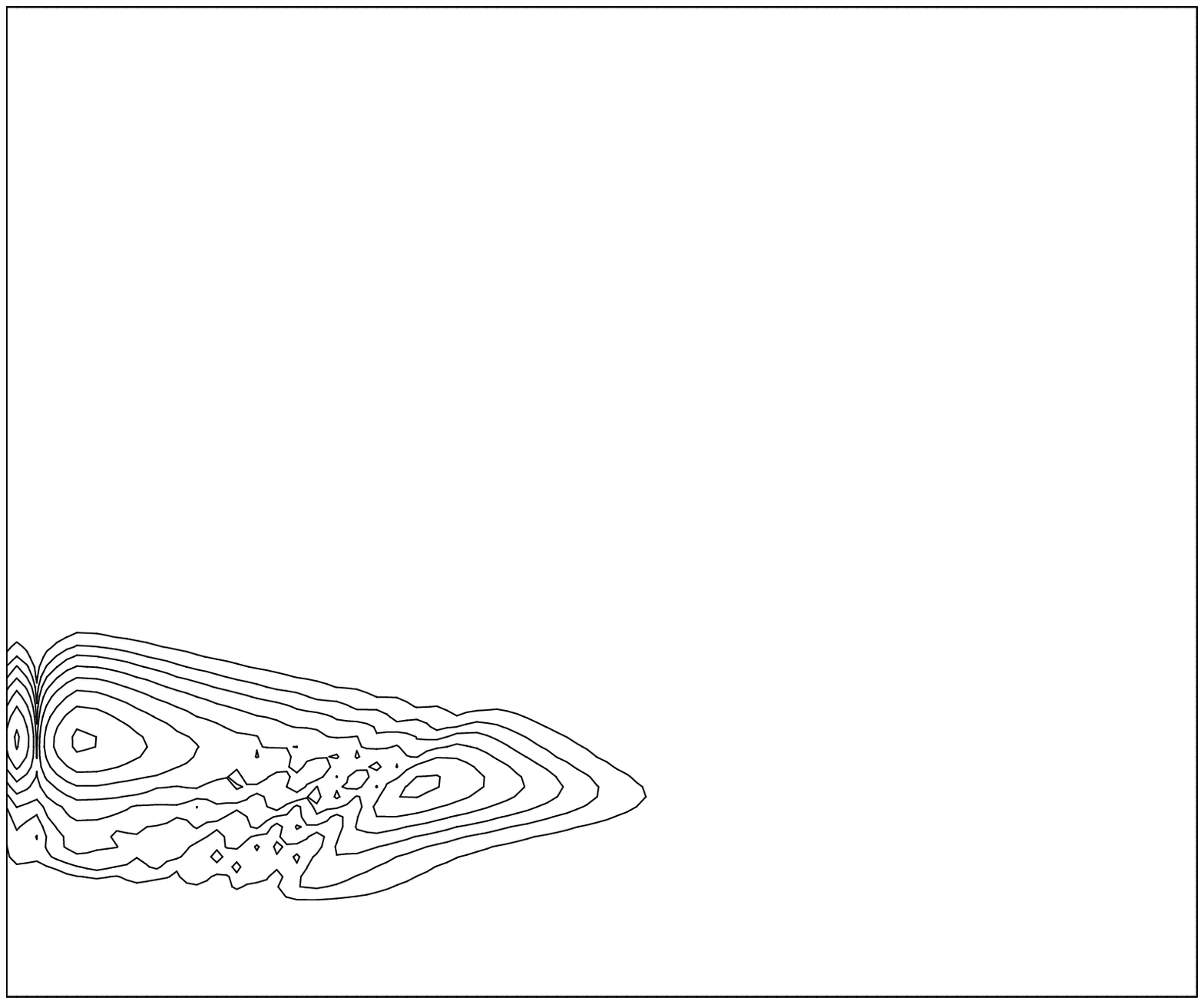}} & &
\resizebox{!}{36mm}{\includegraphics{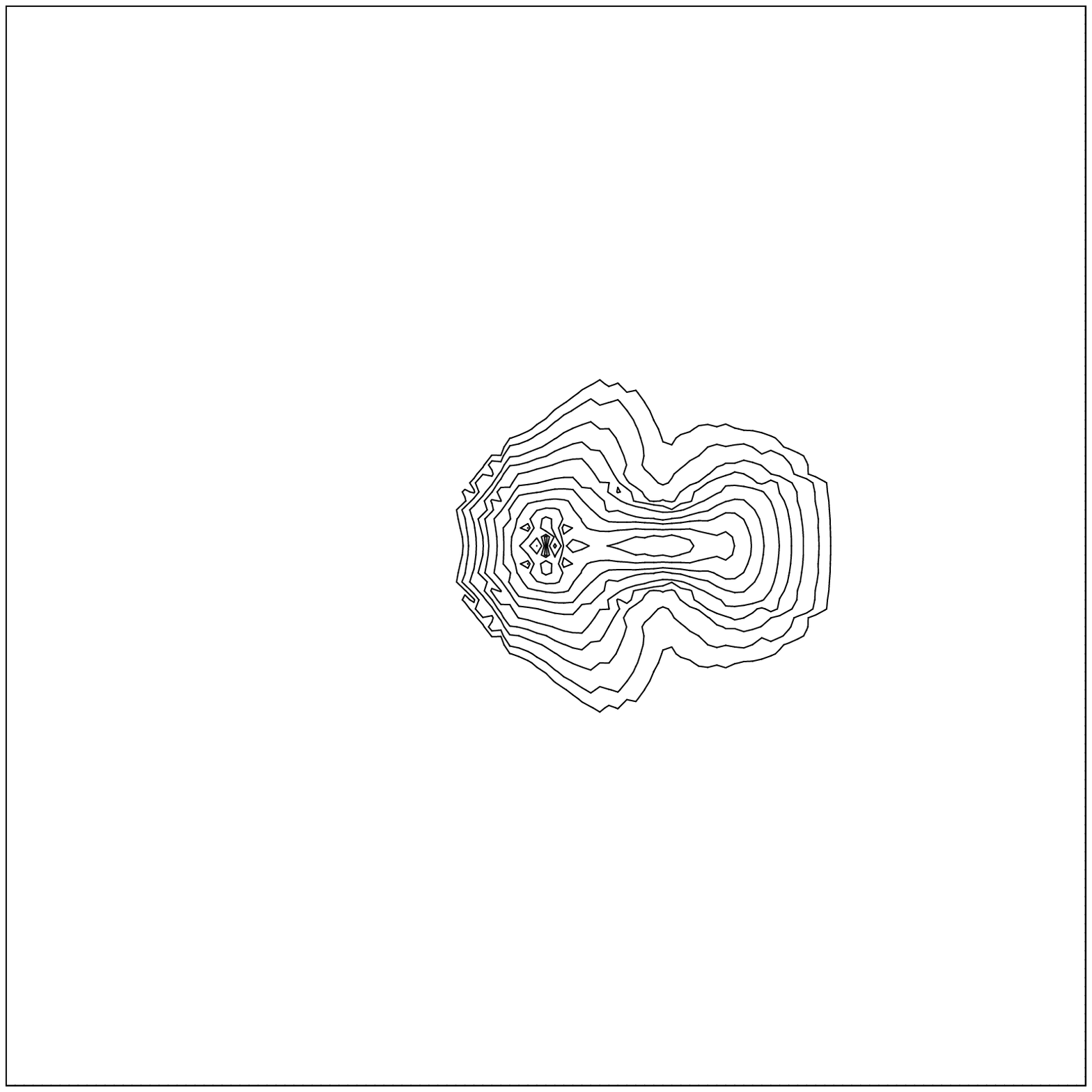}}  \\
\resizebox{!}{36mm}{\includegraphics{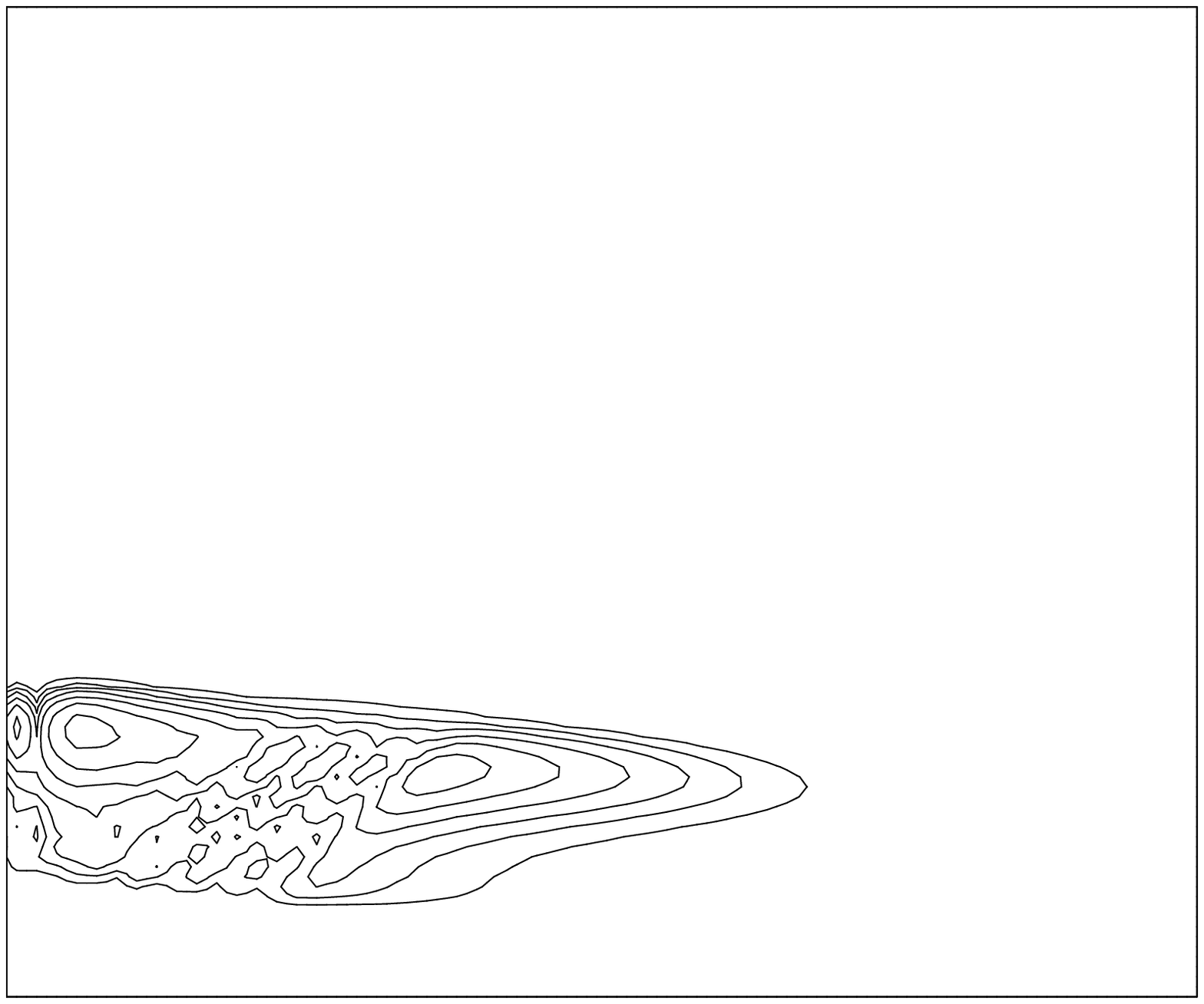}} & &
\resizebox{!}{36mm}{\includegraphics{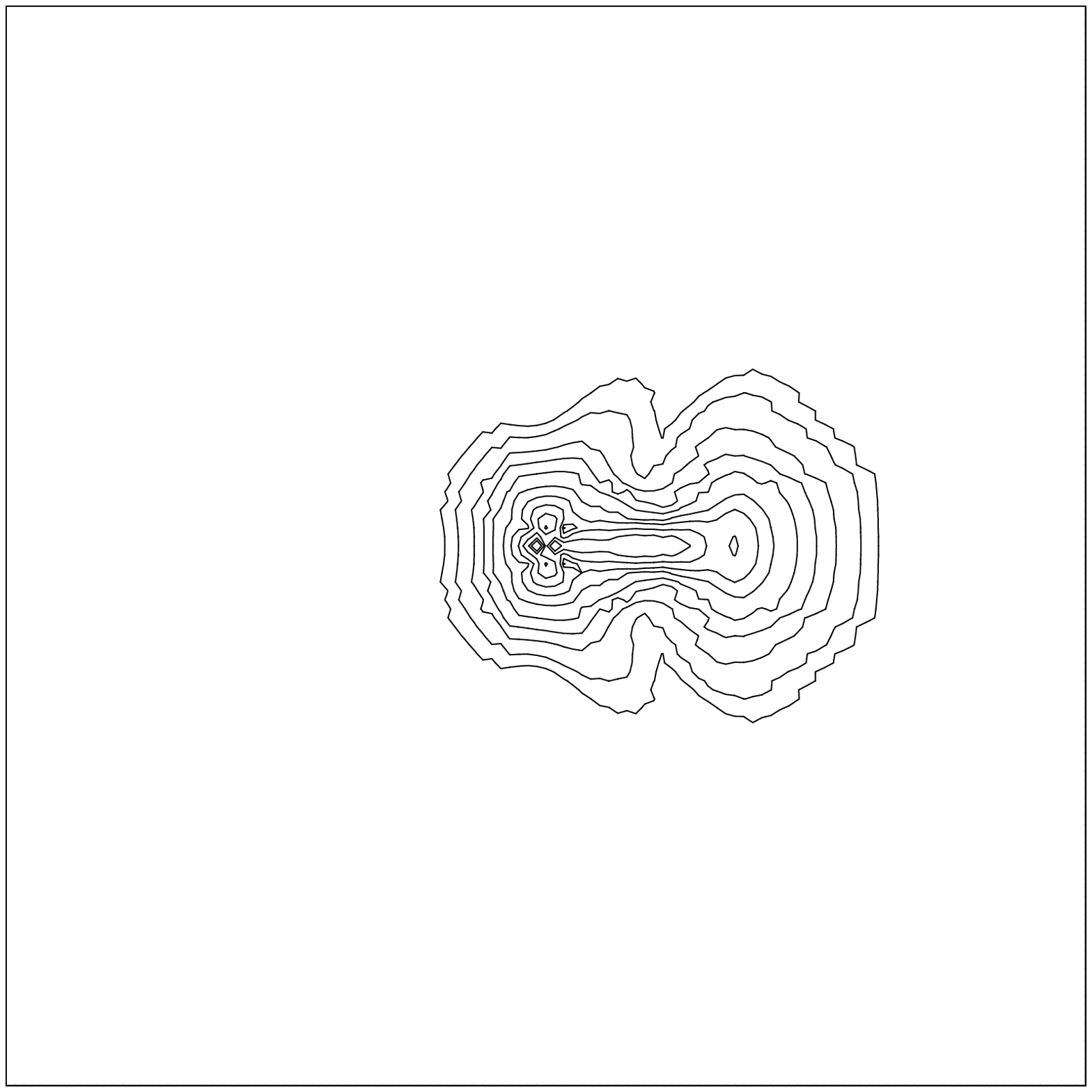}}  \\
\resizebox{!}{36mm}{\includegraphics{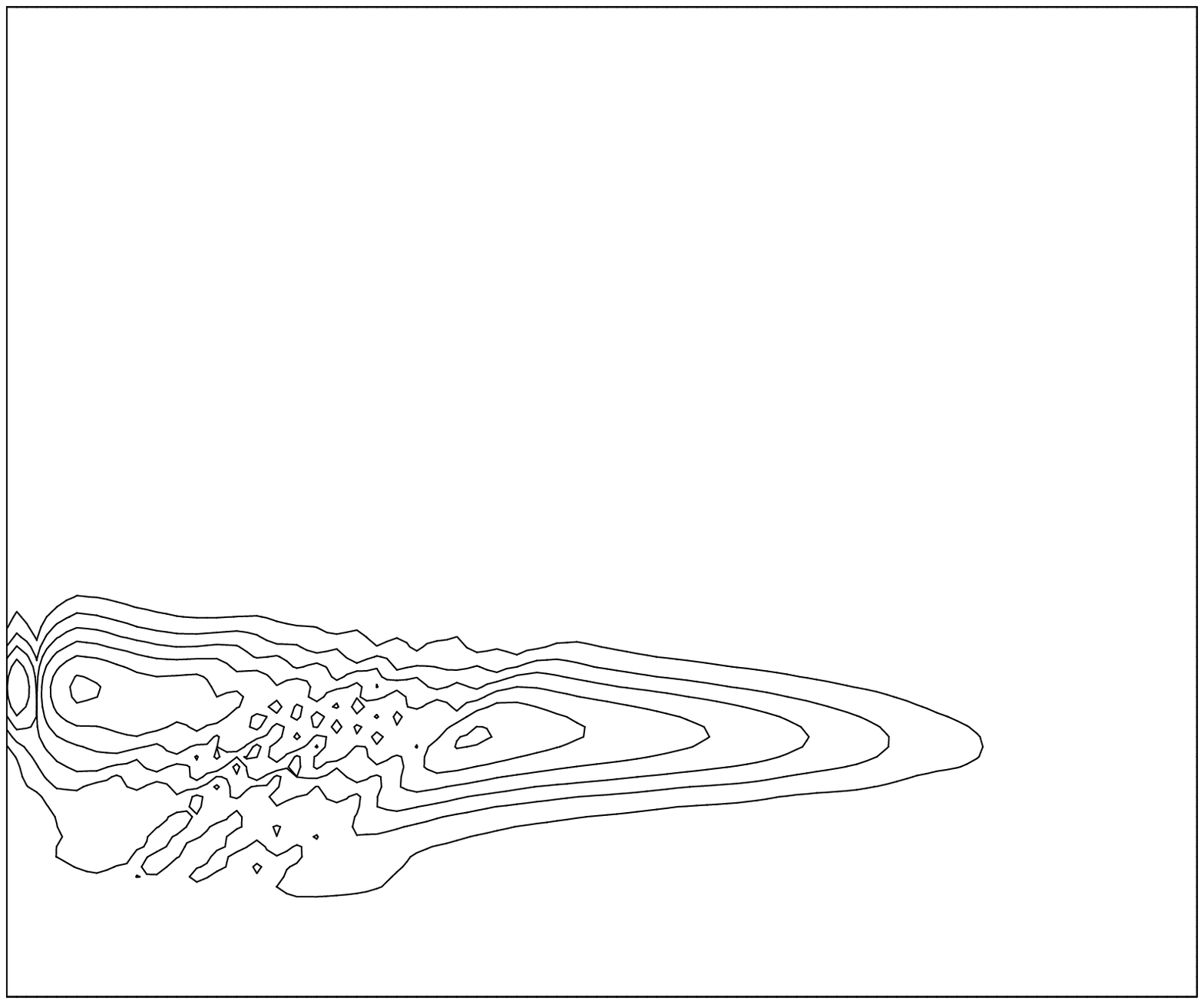}} & &
\resizebox{!}{36mm}{\includegraphics{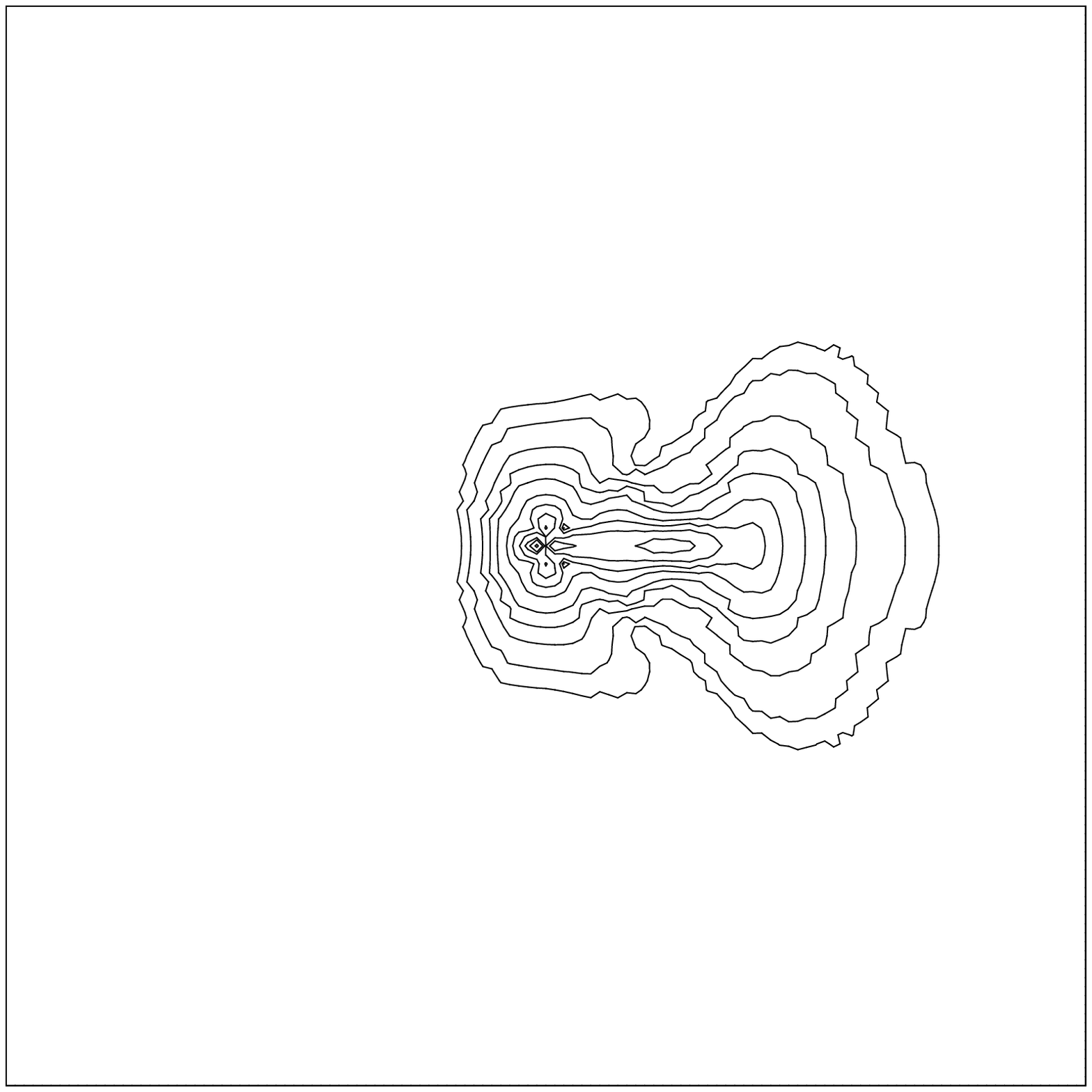}}  \\
\resizebox{!}{36mm}{\includegraphics{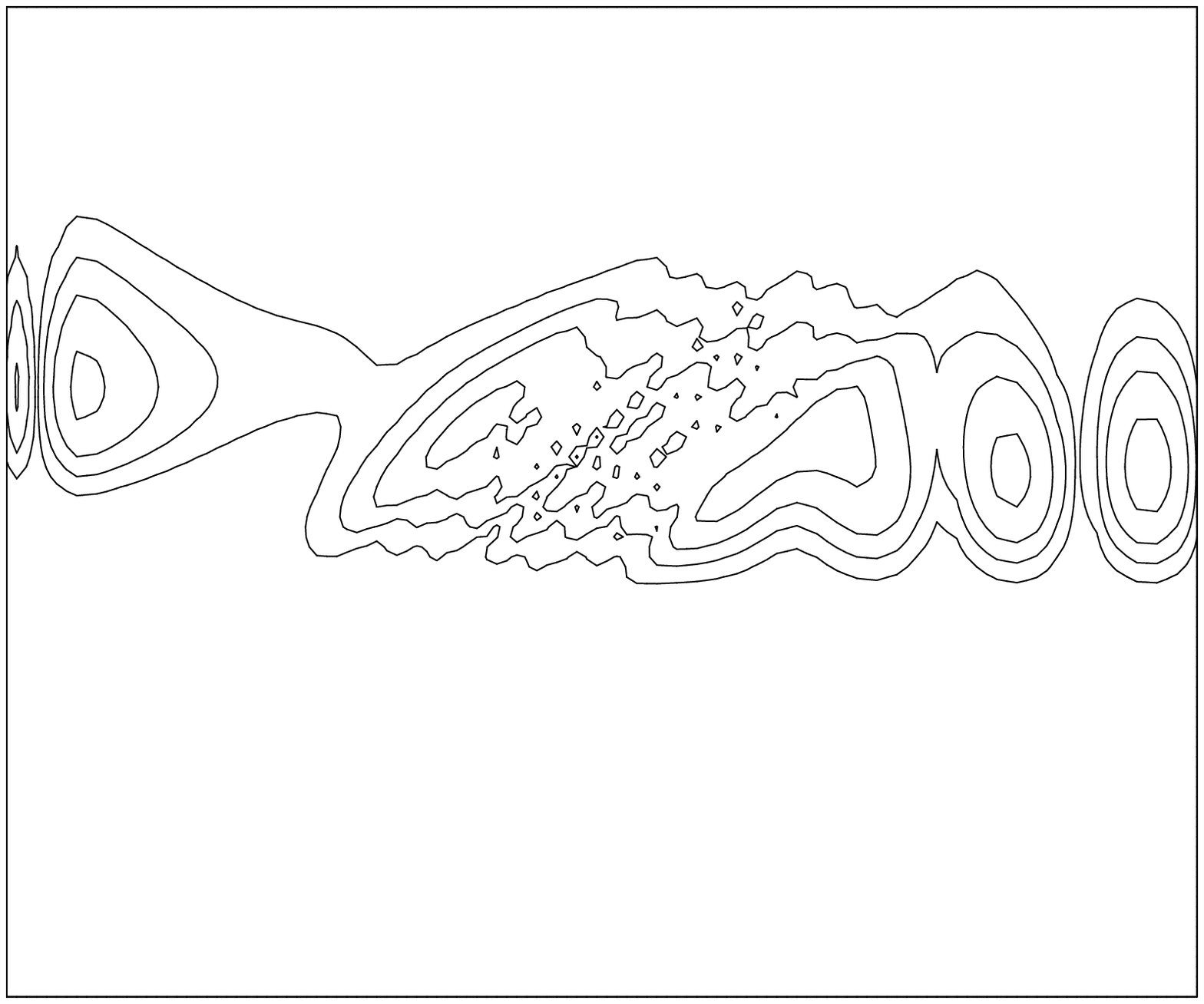}} & &
\resizebox{!}{36mm}{\includegraphics{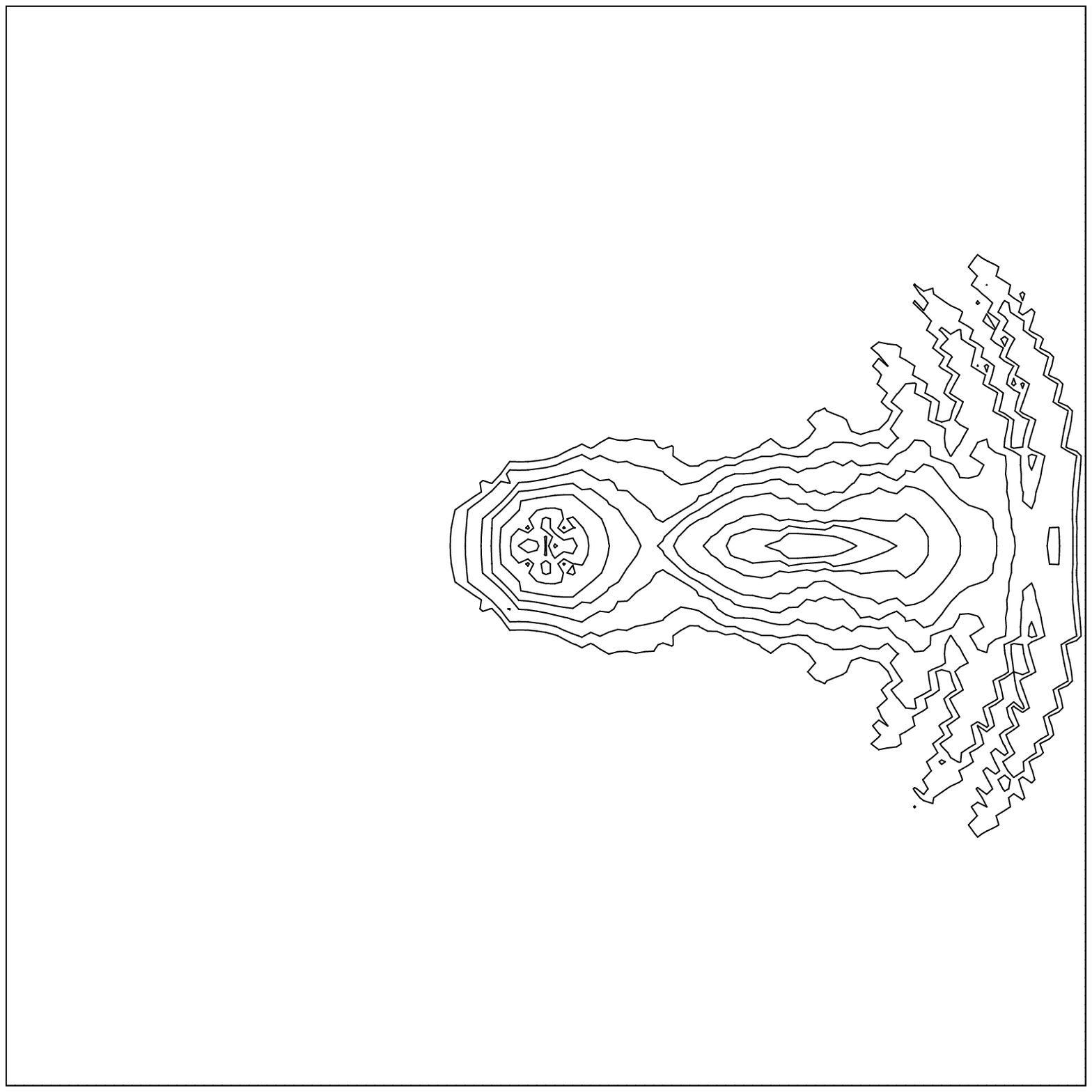}}  \\
\end{tabular}
\end{center}
\caption{The time evolution of the wave packet solution.
In the left panels, the density is plotted as a 
function of $R$ and $r$. The abscissa is the P-T 
distance $R$, and the ordinate is the n-C distance $r$.
In the right panels, the density is plotted as a
function of $r$ and $\theta$ of n-C coordinate ${\bf r}$,
after integrating over $R$.
The initial wave packet is shown in the top panel, and the
final state in the bottom. See text for more detail.
}
\label{fig:1}
\end{figure}

The wave packet is a superposition of the solutions for various 
total energies around $E_0$. It provides us an intuitive picture for the 
reaction dynamics around this energy. In Fig.~\ref{fig:1}, 
we show a time evolution of the wave packet. 
The $K_0$ value is so chosen that the average total
energy $E_0$ is about 38 MeV, which is close to the barrier top energy.
The left-panels show the density distribution integrated over the
angle $\theta$, 
$\rho(R,r,t)=\frac{1}{R^2 r^2}\sum_l \vert u_l(R,r,t) \vert^2$.
Here the abscissa is $r$ and the ordinate is $R$.
The right-panels show the density distribution integrated over $R$,
$\rho(r,\theta,t)=\frac{1}{r^2}\int dR \vert \sum_l u_l(R,r,t)
\left[ \frac{2l+1}{2} \right]^{1/2} P_l(\cos\theta) \vert^2$.
Here the abscissa is $x=r\cos\theta$ and the ordinate is $y=r\sin\theta$.
The center corresponds to the center-of-mass of the projectile, and the
$x$ direction is parallel to the ${\bf R}$ vector. In the figure, the target
nucleus approaches to the center from the right, then returns back.
The time evolves from the top to the bottom.

Let us first look at the five panels in the left.
The top-left panel shows the initial wave packet. There is a nodal
structure of the 2s orbital in the $r$ direction.
The distribution extends to large $r$ reflecting the halo structure.
In the $R$ direction, the wave packet shows the Gaussian
distribution around $R_0$. 
As the time evolves, the wave packet moves towards smaller $R$. 
The third (middle) panel roughly corresponds to the classical turning point.
Some components of the wave packet pass through the C-T 
Coulomb barrier, while some are reflected 
by the barrier. The components going beyond the barrier are
absorbed by the imaginary C-T potential and disappear.
After the wave packet reaches the closest point,
a flow to larger $r$ is seen.
This corresponds to the breakup process.
Performing the calculation with the n-T potential being switched off,
we have confirmed that this is induced by the C-T Coulomb force.
One can also see 
the wave packet component in the diagonal $(R \simeq r)$ direction. 
This is a transfer component. By taking account of high partial waves
($l_{max}=70$), one can describe the transfer processes 
in the Jacobi coordinate system of the incident channel. 
A nodal structure near the right end in the bottom-left panel
is due to a spurious reflection at the boundary in the $r$ coordinate.

The five panels in the right-hand side are useful to see behaviors
of the halo neutron. The initial distribution is spherical since 
we assume $2s$ orbital in $^{11}$Be. Because of the acceleration of the
$^{10}$Be core induced by the target Coulomb field, the halo neutron
appears to be accelerated to the right (towards the target).
Note that the center in these figures corresponds to the center of
mass of the projectile.
This acceleration induces the Coulomb breakup.
As the target nucleus gets closer, one can also see the formation of
molecule-like orbital in the third (middle) panel. In the bottom
panel, the breakup component shows the fan-shaped, multiple waves. This is
again caused by the spurious reflections at the boundary in $r$ coordinate.
The transfer component is also seen as an oval distribution which
locates its center at the average position of the target nucleus.

\section{FUSION PROBABILITY}

To obtain the fusion probability for a fixed incident energy from
the wave packet solution, we apply the energy projection procedure. 
For a wave packet state $\psi_a$, we define the energy distribution 
function $W_a(E)$ as
\begin{equation}
\label{W(E)}
W_a(E) = \langle \psi_a \vert \delta(E-H) \vert \psi_a \rangle,
\end{equation}
where $H$ is the three-body Hamiltonian in Eq.~(\ref{TDSE}).
We evaluate Eq.~(\ref{W(E)}) for the initial ($a=i$)
and the final wave packet ($a=f$).
The fusion probability is then obtained as a fraction of the energy
distribution that was absorbed by the C-T absorptive potential.
\begin{equation}
P_{\rm fusion}(E) = \frac{W_i(E) - W_f(E)}{W_i(E)}.
\end{equation}
In calculating the energy distribution function $W_a(E)$, one may also 
employ the time evolution of the wave packet. Fourier transforming
the delta function, one obtains
\begin{equation}
W_a(E) = \frac{1}{\pi\hbar}{\rm Re}
\int_0^{\infty} dt e^{iEt/\hbar}
\langle \psi_a \left( -\frac{t}{2} \right) \vert
\psi_a \left( \frac{t}{2} \right) \rangle ,
\end{equation}
where $\psi_a(t)$ is the solution of the time-dependent Schr\"odinger
equation (\ref{TDSE}) with the initial condition $\psi_a(0)=\psi_a$.

\begin{figure}[tb]
\begin{minipage}[t]{0.47\textwidth}
\includegraphics[width=\textwidth]{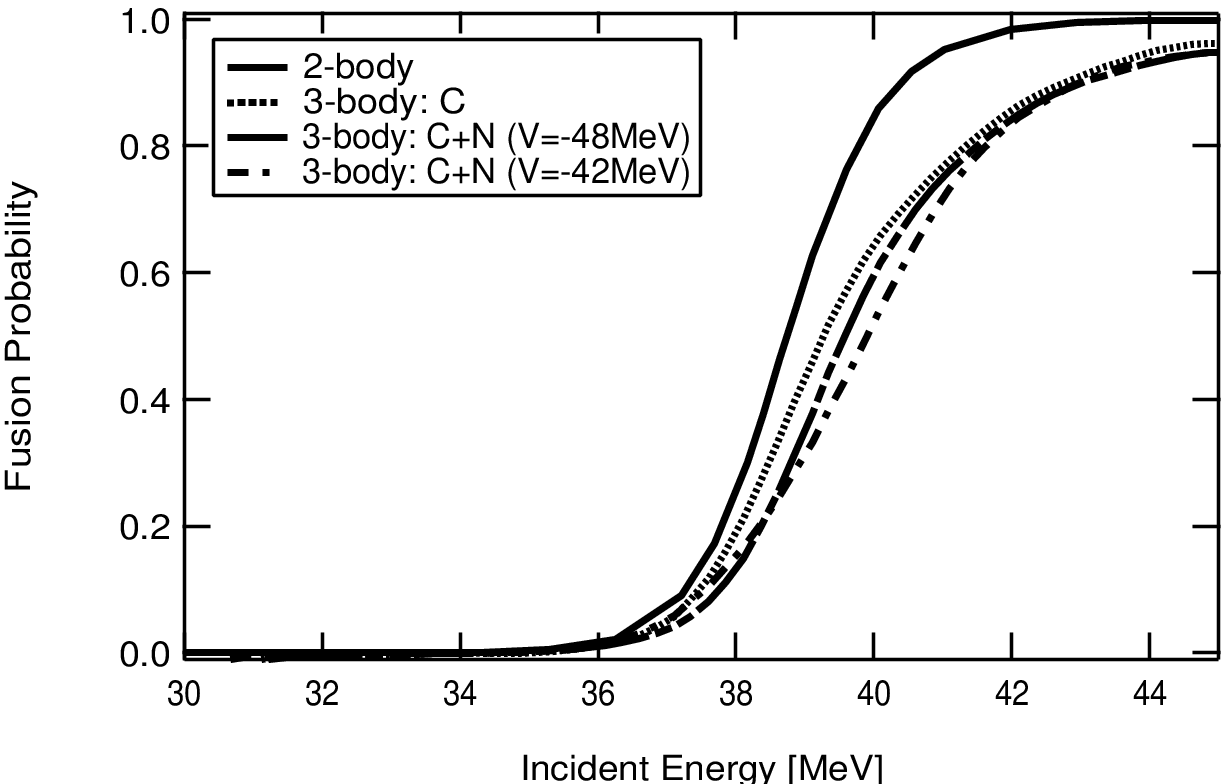}
\caption{Fusion probability as a function of the projectile-target
relative incident energy for the case of zero total angular momentum.
Results of various three-body calculations are compared. See text for 
the details.
}
\label{fig:2}
\end{minipage}
\hfill
\begin{minipage}[t]{0.47\textwidth}
\includegraphics[width=\textwidth]{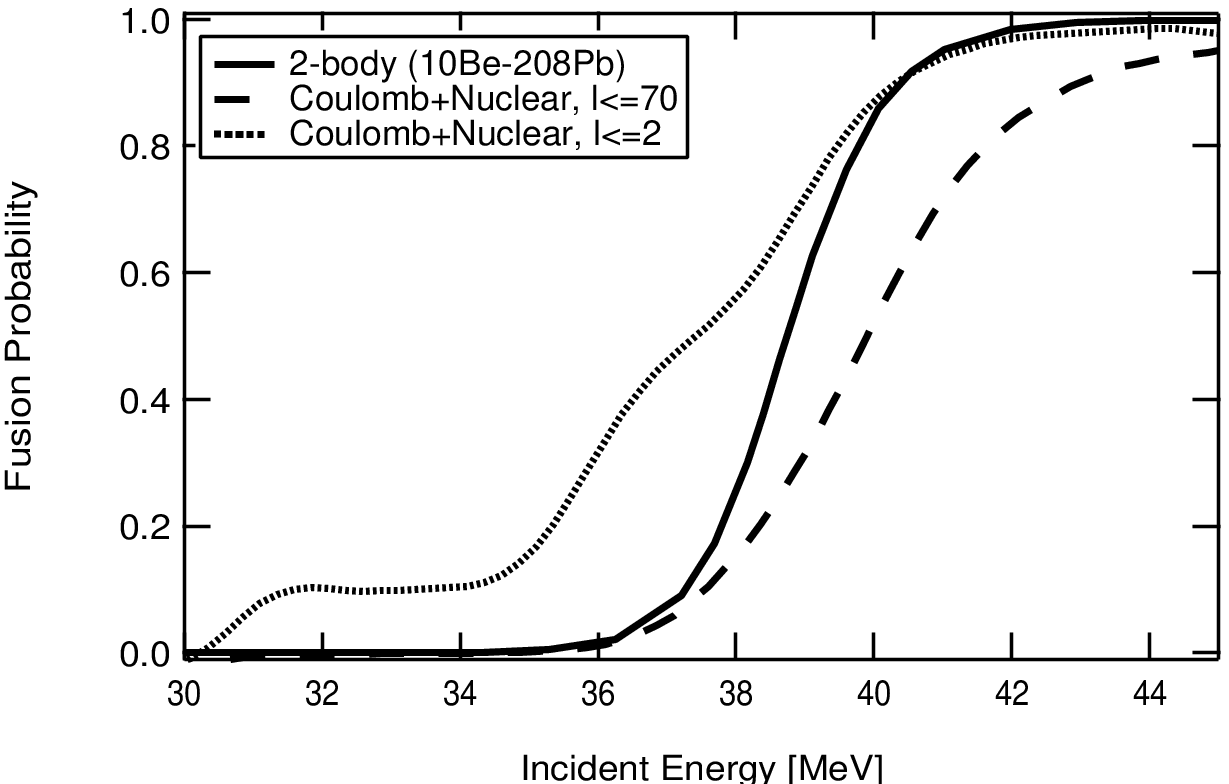}
\caption{
Fusion probability as a function of the projectle-target relative
incident energy. Dependence on the partial wave truncation is shown.
See text for the details.
}
\label{fig:3}
\end{minipage}
\end{figure}

Let us elucidate a basic picture of the reaction dynamics
from the calculated fusion probabilities. 
Figure \ref{fig:2} shows the fusion probability of central collision 
(zero total angular momentum) as a function of the P-T relative 
incident energy. The fusion probabilities of three-body
calculations are compared with that of the
two-body $^{10}$Be-$^{208}$Pb reaction without a halo neutron.
The three-body calculations have been carried out with different
n-T potentials: (i) $V_{nT}=0$ (dotted curve), (ii) $V_{nT}=-48$ MeV
(dot-dashed), and (iii) $V_{nT}=-42$ MeV (dot-dot-dashed).
The Fig.~\ref{fig:1} corresponds to the case (iii), where the
substantial transfer component is seen. There is small transfer
component in the case of (ii).

As is seen from the figure, the fusion probabilities in the three-body 
calculation are always smaller than that of the two-body calculation. 
In other words, the fusion probability is suppressed by the presence
of the halo neutron. Furthermore, this suppression is caused principally
by the C-T Coulomb interaction. The n-T potential plays a minor role
in this fusion reaction.

We could imagine two possible mechanisms that are responsible for
the fusion suppression. The first one is a spectator role of the 
halo neutron. As is seen in Fig.~\ref{fig:1}, 
the halo neutron is emitted to the forwared direction. This indicates
that the halo neutron proceeds straight keeping the incident velocity,
while the core nucleus is reflected by the target Coulomb field. 
In this spectator picture, the relative energy between the core and 
the target nuclei is effectively smaller than the projectile-target
relative energy. This mechanism, which we discussed previously\cite{Yabana97},
is expected to shift the energy dependence of the fusion probability
by a constant amount. The calculated fusion probability, however, 
does not show such a simple shift in energy but shows the stronger
suppression above the barrier. The second mechanism that could 
explain this property is the excitation of the n-C motion. The 
Coulomb interaction between the core and target nuclei induces the 
internal excitation of the projectile. The excitation energy will
distribute in a certain energy region with a fluctuation.  The energy 
conservation requires that the P-T relative energy decreases to 
cancel the projectile excitation energy. This results in the decrease
of the effective incident energy with the fluctuation. The suppression
of the fusion probability persists even high above the barrier, 
if there is a component of high excitation energy in the 
n-C motion. Of course the above two mechanisms are not independent 
but mutually related. More analyses are necessary to obtain clear 
understanding.

Recently, three-body calculations for the fusion reaction of the same 
system have been reported. These studies, however, report the conclusion
opposite to ours, namely the strong enhancement of the fusion probability 
around the barrier energy\cite{Hagino00,Diaz-Torres02}.
Let us next consider a reason for this discrepancy. In these 
three-body calculations, the continuum states are discretized in 
the momentum space. The partial wave expansion of the n-C motion is 
truncated at very low angular momentum, a few units in the $\hbar$. 
We believe that this truncation is the origin of the discrepancy, 
causing even an opposite conclusion. Figure \ref{fig:3} shows
the fusion probability calculated with $l_{max}=70$ (dashed curve)
and that with $l_{max}=2$ (dotted curve).
As is seen clearly, the calculation taking only low partial waves 
gives strong enhancement of the fusion probability at low incident
energy. However, increasing the number of partial waves, 
this enhancement completely disappears and the fusion probability 
is actually suppressed in the entire energy region. 
Therefore, it is highly probable that the fusion enhancement obtained in 
Refs.\cite{Hagino00,Diaz-Torres02} is a false account derived from the
calculations which did not converge with respect to the partial waves.

We have found that the partial wave expansion up to small $l$-value
gives a qualitatively correct answer when one includes only Coulomb
breakup mechanism switching-off the n-T nuclear potential. 
Therefore, it is the n-T nuclear potential that does not allow the 
partial wave truncation. A similar behavior is reported in
Ref.~\cite{Esbensen01} which points out that the inclusion of the 
high partial waves is necessary to obtain converged breakup cross 
section at low incident energies.

\section{OUTLOOK}

To calculate the fusion cross section, we need to make calculations for
reactions at finite total angular momenta. This is under progress in
the body-fixed frame descrption, where the computational cost scales 
linearly to the total angular momentum.

For the three-body direct reactions, coupled-channel frameworks 
discretizing the continuum channels in momentum space have been
developed and successfully applied. The present real-space
approach for the three-body problem is more straightforward but 
computationally expensive. As the rapid progresses of the computational
resources, we expect the real-space approach will be more powerful
in the future. We would like also to mention that we have recently
developed an alternative real-space approach for the breakup reaction 
employing the absorbing boundary condition\cite{Ueda02}.

\end{document}